\documentclass[aps,twocolumn,showpacs,prl,amsmath,amssymb]{revtex4-1}
\newcommand{\lsim} 
 {\ \raise.35ex\hbox{$<$}\kern-0.75em\lower.5ex\hbox{$\sim$}\ }
\newcommand{\gsim}
 {\ \raise.35ex\hbox{$>$}\kern-0.75em\lower.5ex\hbox{$\sim$}\ }

\usepackage{graphicx}
\usepackage{dcolumn}
\usepackage{bm}
\usepackage{amsmath}
\usepackage{times}
\usepackage[usenames]{color}
\usepackage{natbib}
\usepackage{ulem}
\begin{document}
\title{Optical suppression of electron motion in low-dimensional correlated electron system}  
\author{Atsushi Ono}
\author{Hiroshi Hashimoto}
\author{Sumio Ishihara}
\affiliation{Department of Physics, Tohoku University, Sendai 980-8578, Japan}
\date{\today}
\begin{abstract}  
Suppression of electron motion under an alternating current (AC) electric field is examined in a one-dimensional Hubbard model.  
Utilizing three complementary calculation methods, 
it is found unambiguously that magnitudes of the kinetic-energy suppressions 
are influenced sensitively by the Coulomb interaction as well as the electron density. 
The phase and frequency in the AC field do not bring about major effects. 
The results are interpreted as a combined effect of the Coulomb interaction and the AC field, and 
provide a guiding principle for the photocontrol of correlated electron motion. 
\end{abstract}


\maketitle
\narrowtext



%
%

%




The ultrafast control of electronic states using light has been a challenging and exciting topic in the  field of condensed matter physics for the past several decades~\cite{koshihara,basov,aoki}. 
Recent significant progress in ultrafast optical techniques, x-ray laser facilities, and computational algorithms has accelerated developments concerning fundamental research into electron dynamics. 
One of the attractive topics in this research field is the exploration of photoinduced hidden states~\cite {ichikawa,stojchevska,kaiser,matsuzaki,zhang,moussa}. 
Several transient electronic and structural hidden states, which do not appear under conditions of thermal equilibrium, have been discovered in correlated electron materials owing to their complex degrees of freedoms. 

Another vital and desirable target in this field is establishment of methods to control the electronic parameters of a solid. Once such methods are settled, a core procedure for manipulating the electronic states of matter and their functionalities using light will be obtained. 
The theoretical proposals of the photoinduced sign changes in the electronic interactions and the magnetic exchange interactions~\cite{tsuji,mentink} may yield new routes towards optical manipulation of magnetism and superconductivity. 
%

The suppression of the electron motion induced by an intense alternating current (AC) field  
is a prototypical example of the light-control of electronic parameters.  
The so-called ``dynamical localization" (DL) phenomenon was predicted for the non-interacting charged particle systems, where the hopping integral $t$ is multiplied by the zeroth-order Bessel function~\cite{dunlap,grossmann,kayanuma}. 
The sign and magnitude of the effective $t$ are expected to be adjusted by 
varying the light parameters. 
This approach has been recognized as a successful strategy for controlling correlated electron materials; competitions and cooperations between the kinetic-energy suppression and the Coulomb interaction energy provide the phase instabilities and the novel photoinduced phenomena~\cite{eckardt,lignier,nishioka,yonemitsu,ishikawa,naitoh}.   
However, despite the recent intensive researches, the kinetic-energy suppression phenomenon itself has  been addressed within the original framework of the non-interacting electron system. 
%
%

\begin{figure}[t]
\begin{center}
\includegraphics[width=0.9\columnwidth, clip]{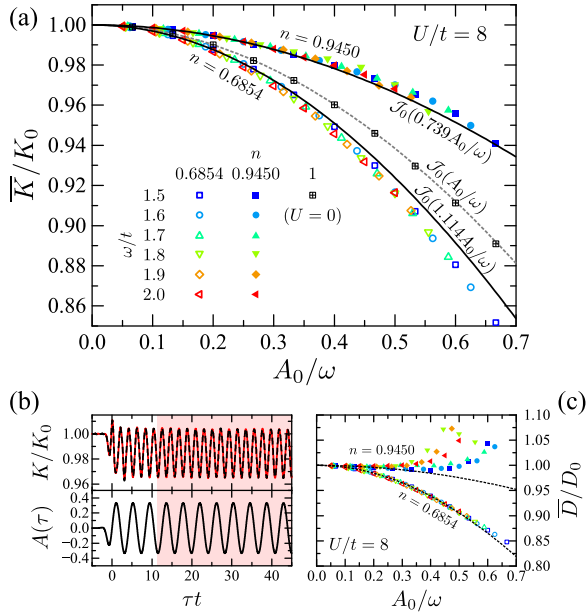}
\end{center}
\caption{(Color online) 
Numerical results yielded by iTEBD method at $U/t=8$.  
(a) Normalized time-averaged kinetic energy $\overline{K}/K_0$ at $n=0.6854$ and  $0.945$. 
The bold lines indicate least-square fitting using Eq.~(\ref{eq:bessel}) and 
the dotted line corresponds to $\mathcal{J}_0(A_0/\omega)$. 
$\overline{K}$ is obtained from $K(\tau)$ in $45/t-16\pi/\omega<\tau<45/t$. 
Squares with crosses are yielded by iTEBD at $U/t=0$ and $n=1$. 
%
(b) Time profiles of vector potential $A(\tau)$ and normalized kinetic energy $K(\tau)/K_0$. 
The results for $\chi=50$ and $100$ are represented by dotted and bold lines, respectively. 
The shaded area represents the time domain in which  
$\overline{K}$ in (a) is calculated. 
Parameter values are chosen to be $\omega/t=1.5$ and $\tau_{\rm p}=1/t$. 
(c) Normalized time-averaged double occupancy $\overline{D}/D_0$, at $n=0.6854$ and $0.945$, 
where $D_0=\langle 0 |\sum_i n_{i \uparrow} n_{i \downarrow} |0 \rangle$. 
The dotted lines are guides for the eye. 
}
\label{fig:ek}
\end{figure}

In this Letter, we perform a semi-qualitative investigation of the manner in which the electron correlation effect influences the photoinduced electron motion suppression in a correlated electron system. 
As a prototypical correlated electron model, we analyze a one-dimensional Hubbard model under an AC electric field. 
Unambiguous results are obtained by using three complementary methods: the infinite time-evolving block decimation (iTEBD) algorithm, 
the Floquet theory combined with the exact diagonalization (ED) method, termed ``Floquet+ED", and the perturbation method. 
The kinetic energy is obtained for wide parameter regions of the on-site Coulomb interaction $U$, the electron density $n$, and the amplitude $A_0$, frequency $\omega$,  and phase $\phi$ of the AC field. 
The kinetic-energy suppressions vary sensitively in response to both $n$ and $U/t$. 
Typical examples of the normalized time-averaged kinetic energy 
are presented in Fig.~\ref{fig:ek}(a).
The present results indicate that the Coulomb interaction does not only induce the system instabilities 
in cooperation with the photoinduced kinetic-energy suppression, but also influences the suppression phenomenon itself. 

The Hubbard model in the one-dimensional lattice analyzed in the present paper is defined as 
\begin{align}
{\cal H}=&- \sum_{i \sigma} \left (t c_{i \sigma}^\dagger c_{i+1 \sigma}+ \text{H.c.} \right ) \nonumber \\
&+U\sum_i \left ( n_{i \uparrow}-\frac{1}{2}\right ) \left (n_{i \downarrow}-\frac{1}{2} \right), 
\label{eq:model}
\end{align}
where $c_{i \sigma}^\dagger$ ($c_{i \sigma}$) is the creation (annihilation) operator for an electron at site $i$ with spin $\sigma$, $n_{i \sigma}=c_{i \sigma}^\dagger c_{i \sigma}$ is the number operator, 
$t$ is the hopping integral between the nearest neighboring sites, and $U$ is the on-site Coulomb interaction. 
The first and second terms are denoted as $ {\cal H}_{\rm kin}$ and ${\cal H}_{\rm int}$, respectively. 
The time-dependent external field is introduced in $ {\cal H}_{\rm kin}$ as the Peierls phase as   
$t \rightarrow t e^{-i A(\tau)}$ where $A(\tau)$ is the vector potential at time $\tau$. 
The Hubbard Hamiltonian in which $A(\tau)$ is introduced is denoted as $ {\cal H}_A$. 
The light velocity, lattice constant, elementary charge, and Planck constant are set to one, and the  Coulomb gauge is adopted. 
The semi-infinite AC field is applied along the chain, having the form 
\begin{align}
A(\tau)=
\begin{cases}
\displaystyle      (A_0/\omega) e^{-\tau^2/(2\tau_{\rm p}^2)} \sin(\omega \tau) &  \  (\tau < 0) , \\
\displaystyle    (A_0/\omega) \sin(\omega \tau)   & \   (\tau \geq 0) , 
\end{cases}
\label{eq:ac}
\end{align}
where we chose $\tau_{\rm p}=1/t$ for the numerical calculations. 
The electron density is $n=N/L$, where $N$ and $L$ are the electron and site numbers, respectively, and 
$n=1$ corresponds to the half filling.

The electronic states under the AC field are analyzed by utilizing three complementary methods. 
First, we introduce the results obtained using the iTEBD methods~\cite{vidal,ors,takayoshi,ono},
which are known as an efficient simulation algorithm for quantum many-body systems in the thermodynamic limit. 
The wave function is represented in the matrix-product form~\cite{vidal,ono} 
with the matrix dimension $\chi$.
The ground state $| 0\rangle $ is calculated by using the infinite density-matrix renormalization group method~\cite{mcculloch}. 
The time-evolved states are obtained from  
$|\Psi (\tau) \rangle \propto \prod \exp(-i \delta \tau {\cal H}) |0 \rangle$, 
with
a small time difference $\delta \tau$. 
For most of the numerical calculations presented in this paper, we chose $\chi=100$ and $\delta \tau=0.01/t$. 
Note that the truncation error in the ground state at $U/t=8$, 
$n=0.945$, and $\chi=100$ is less than $10^{-5}$. 
The deviations of $n$ in the time evolved-state are less than $10^{-4}$ until $\tau=50/t$. 
%

\begin{figure}[t]
\begin{center}
\includegraphics[width=0.8\columnwidth, clip]{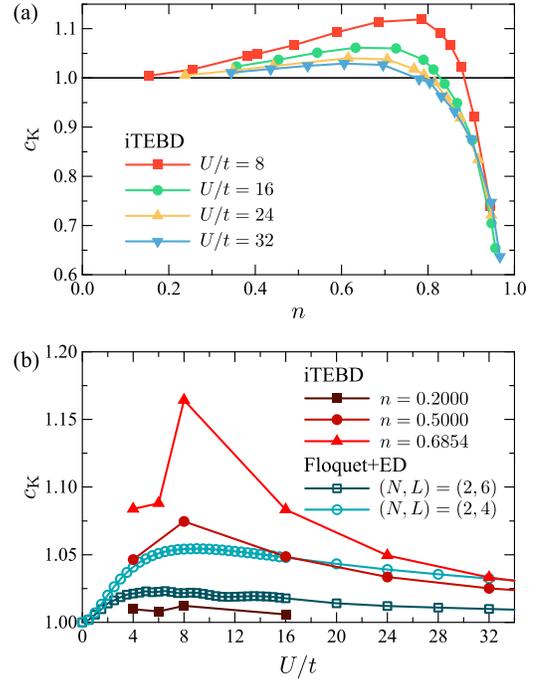}
\end{center}
\caption{(Color online) 
(a) Suppression factors $c_{\rm K}$ of the time-averaged kinetic energy under AC field (see Eq.~(\ref{eq:bessel})) as functions of  $n$ for several $U/t$, and  (b) $c_{\rm K}$ as functions of $U/t$ for several $n$. 
The filled and open symbols show data calculated using the iTEBD method and the Floquet+ED method, respectively. 
Parameters are chosen to be $\omega/t=1.5$ and $\chi=100$ for the iTEBD method, and $\omega/t=0.8$ for the Floquet+ED method. The upper limits of the number of Floquet states (see text) are chosen to be $N_{\rm ph}=8$ for $(N, L)=(2, 6)$, and $N_{\rm ph}=16$ for $(N, L)=(2, 4)$. 
}
\label{fig:main}
\end{figure}

Typical time profiles of $A(\tau)$ and the normalized kinetic energy $K(\tau)/K_0$  are shown in Fig.~\ref{fig:ek}(b). 
We define $K(\tau)=\langle \Psi(\tau) | {\cal H}_{\rm kin} |\Psi(\tau) \rangle$, which is  measurable as the total weight of the optical spectra, and $K_0$ is the kinetic energy without the external field.
As the AC field is introduced at approximately $\tau=0$,  $K(\tau)/K_0$ responds rapidly, and 
the system moves into a steady state smoothly. 
The differences between the calculated results for $\chi=50$ and $100$ are less than $0.2\%$ [see the bold and dotted lines in Fig.~\ref{fig:ek}(b)].  
We note that both $K(\tau)$ and $K_0$ are negative, and $K(\tau)/K_0$ measures the absolute value of the kinetic energy. 
%
The time-averaged kinetic energy $\overline{K}$ is calculated from $K(\tau)$ in the stational state indicated in Fig.~\ref{fig:ek}(b). 
The reduction of $\overline{K}/K_0$ is termed the kinetic-energy suppression in the present paper. 
In a similar manner, the time-averaged double occupancy $\overline{D}$ is deduced from the time profiles of   
$D(\tau)=\langle \Psi(\tau) |\sum_i n_{i \uparrow} n_{i \downarrow} |\Psi (\tau) \rangle$. 

The results of $\overline{K}/K_0$ at $n=0.685$ and $0.945$ for $U/t=8$ are presented in Fig.~\ref{fig:ek}(a). 
Each data set for several $\omega$ is scaled on a single curve as a function of $A_0/\omega$, 
and the data within $A_0/\omega \le 0.3$ are well fit by the zeroth-order Bessel function defined as 
\begin{align}
\overline{K}/K_0=\mathcal{J}_0 \left(c_{\rm K} A_0/\omega \right ) . 
\label{eq:bessel}
\end{align} 
Here, a numerical factor $c_{\rm K}$, termed the ``suppression factor", measures  the suppression magnitude and $c_{\rm K}=1$ is expected from the conventional DL theory in a  non-interacting system [see broken line in Fig.~\ref{fig:ek}(a)]. 
Using the least-square fitting, we obtain $c_{\rm K}=1.114$ ($n=0.685$) and $0.739$ ($n=0.945$), within 0.3$\%$; these results are unambiguously larger and smaller than one, respectively. 
We restrict the present iTEBD analyses to $\omega/t \lsim 2$ and $A_0/\omega \lsim 0.5$, where 
accuracy is guaranteed. 
The characteristic oscillations in the Bessel function appearing in $A_0/\omega >2$ are not confirmed.  
Note that the results for larger $\omega$ calculated using the Floquet+ED method are shown below. 
The results of $\overline{D}$ at $n=0.685$ and $0.945$ are shown in Fig.~\ref{fig:ek}(c). 
As $\overline{D}/D_0$ increases at $n=0.945$ for $A_0/\omega \ge 0.35$, which is most likely due to resonant-like transitions, we restrict our analyses to the $A_0/\omega<0.3$ region. 
$\overline{D}$ decreases with increasing $A_0/\omega$ for both $n$ values, 
which means that the two electrons with opposite spins avoid to occupy the same site.
The slope of the curve at $n=0.685$ is steeper than that at $n=0.945$. 
The suppression of $\overline{D}$ by the AC field and the difference between the two results at $n=0.6854$ and $0.945$ are consistent with the reductions in $\overline{K}/K_0$ shown in Fig.~\ref{fig:ek}(a). 
The numerical results obtained using the iTEBD method are summarized in Fig.~\ref{fig:main} (filled symbols). 
The kinetic energy suppression exhibits a sensitive dependence on both $n$ and $U/t$; 
$c_{\rm K}$ is larger than one for $n \le 0.85 $, and is significantly smaller than one for an around a half-filled case. 
The data seem to be extrapolated to $c_{\rm K}=1$ at $U/t=0$, as well as $n=0$, as expected from the standard DL theory.

\begin{figure}[t]
\begin{center}
\includegraphics[width=\columnwidth, clip]{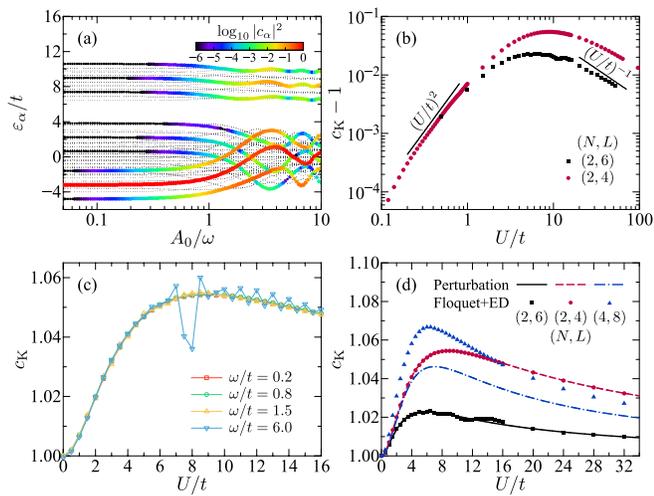}
\end{center}
\caption{(Color online) 
(a)--(c) Numerical results obtained using the Floquet+ED method. 
(a) Floquet quasi-energies $\varepsilon_\alpha$ as functions of $A_0/\omega$. The weights $|c_\alpha|^2$ for each quasi-energy are represented by color. Parameter values are chosen to be $U/t=8$ and $(N, L)=(2, 4)$. 
(b) Logarithmic plots of $c_{\rm K}$ as functions of $U/t$ at $\omega/t=0.8$. 
(c) The $U/t$ dependence of $c_{\rm K}$ for several $\omega$.
We chose $N_{\rm ph}=2$ in (a), and $N_{\rm ph}=8$ for $(N, L)=(2,6)$ and $N_{\rm ph}=16$ for $(N, L)=(2, 4)$ in (b) and (c).   
(d) Numerical results yielded by perturbation method. 
The filled symbols show the results yielded by the Floquet+ED method, for comparison. 
}
\label{fig:floquet}
\end{figure}
%
The $U/t$ dependence of the kinetic energy suppression can be clarified more precisely using the Floquet+ED method. 
Here, we focus on systems far from the half filling ($n\lsim 0.8$). 
Instead of the $A(\tau)$ given in Eq.~(\ref{eq:ac}), $A(\tau)=(A_0/\omega)\sin \omega \tau $ for all time is introduced in the Hubbard Hamiltonian.  
The eigen-value equations for the Floquet states~\cite{shirley,sambe} are given by
\begin{align}
\sum_{m}  H^\omega_{nm} |\phi_\alpha^m\rangle
= \varepsilon_\alpha |\phi_\alpha^n \rangle , 
\label{eq:floquet}
\end{align}
with the Floquiet Hamiltonian being expressed as 
\begin{align}
 H^\omega_{nm}= H_{n-m}-m \omega \delta_{mn} . 
\label{eq:floquetH}
\end{align} 
We define that $ H_m$ and $|\phi_\alpha^m\rangle$ are the $m$-th Fourier components of the time-dependent Hamiltonian ${\cal H}_A$ and  the $\alpha$-th Floquet state $|\phi_\alpha (\tau) \rangle$, respectively, 
and $\varepsilon_\alpha$ is the $\alpha$-th Floquet quasi-energy. 
The wave function at time $\tau$ is given by 
$|\Psi(\tau)\rangle=\sum_\alpha c_\alpha e^{- i \varepsilon_\alpha \tau} |\phi_\alpha (\tau) \rangle$  
with $c_\alpha= \langle \phi_\alpha^{n=0} |0 \rangle$, where $|0 \rangle$ is the ground state of ${\cal H}$. 
%
%
Equation~(\ref{eq:floquet}) is solved in $L$-site clusters, where the number of the Fourier components is truncated to 2$N_{\rm ph}+1$. 
%
The quasi-energies $\varepsilon_\alpha$ and 
the corresponding weights $|c_\alpha|^2$ are shown in Fig.~\ref{fig:floquet}(a).  
For $A_0/\omega<0.5$, one Floquet state is dominant with the weights for the other states being less than 10$^{-3}$. 
In the calculations of $|\Psi (\tau) \rangle $, 
all Floquet states are considered when the Floquet+ED method is employed, and 
the dominant state is considered when the perturbation method is employed. 
Convergences of the results with respect to $N_{\rm ph}$ are confirmed.

The suppression coefficients obtained using the Floquet+ED methods are plotted in Fig.~\ref{fig:main}(b) (open symbols).  
The $U/t$ dependence of $c_{\rm K}$ reproduces the results yielded by the iTEBD methods semi-qualitatively. 
The numerical data are smoothly connected from the weak to strong coupling regimes, and approach  $c_{\rm K}=1$ at $U/t=0$. 
As can be seen from the logarithmic plots shown in Fig.~\ref{fig:floquet}(b), 
$c_{\rm K}$ varies with $(U/t)^2$ and $(U/t)^{-1}$ in the weak and strong coupling limits, respectively. 
This method is also advantageous as it can elucidate the detailed $\omega$ dependence of $c_{\rm K}$ 
[see in Fig.~\ref{fig:floquet}(c)].
The data for a wide frequency range ($0.2 \le \omega/t \le 6$) are well scaled by a single curve.  
That is, the $c_{\rm K}-U/t$ curve does not depend on $\omega$ within these data sets, 
in which the two frequency limits, $\omega \ll U, t $ and $\omega \gg U, t$, are incorporated. 
A dip structure at approximately $U/t=8$ for $\omega/t=6$ is most likely due to the resonant transitions.
%

The physical interpretations of the results presented above are obtained using the perturbation method with respect to the AC field. 
The Floquet Hamiltonian in Eq.~(\ref{eq:floquetH}) is separated into two components, such that  
$H^\omega_{nm}=(V_0)_{nm}+(V_A)_{nm}$, 
where $(V_0)_{nm}$ is defined as $ H^\omega_{nm}$ with $A_0=0$, and $(V_A)_{nm}$ is the remaining component which is treated as the perturbational term. 
%
%
%
The explicit form of $c_{\rm K}$ calculated up to the second-order perturbation is given in the Supplemental Material (SM)~\cite{sm}, and is evaluated numerically using the ED method in finite size clusters.
As shown in Fig.~\ref{fig:floquet}(d), which also shows the data obtained using the Floquet+ED method, 
the results yielded by the two methods almost coincide with each other, although some differences exist in the results for $(N, L)=(4,8)$. 
The good agreement between the two different methods 
can also be seen in the results for $\overline{D}$ (not shown). 

It is confirmed numerically that, of the several terms in Eq.~(S.9) in SM, the second term is dominant, as 
\begin{align}
c_{\rm K}^2 \sim 1+2 \sum_{i (\ne 0)}
\frac{|\langle i| {\cal H}_{\rm kin}|0 \rangle|^2}{K_0(E_0-E_i)} , 
\label{eq:pert}
\end{align}
where $|i \rangle$ is eigen state of ${\cal H}$ with energy $E_i$. 
This equation implies that  $c_{\rm K}$ is greater than one within the present approximation. 
We note that $\langle i |{\cal H}_{\rm kin}|0 \rangle=0$ in the non-interacting system, as $|i \rangle$ and $|0 \rangle$ are the eigen states of ${\cal H}_{\rm kin}$. 
The direct calculations performed using the ED method demonstrate that 
the numerator (denominator) in the second term in Eq.~(\ref{eq:pert}) governs the $U/t$ dependence of $c_{\rm K}$ in the region of $U/t \lsim 5$ ($U/t\gsim 5$). 
%
%
The deviation of $c_{\rm K}$ from one 
is interpreted as being attributable to
the electron scattering near the Fermi level, whereas that in the strong coupling regime is due to the scattering between the remnants of the lower and upper Hubbard bands with the energy differences of  the order of $U$.  
In other word, the electron redistribution to the higher energy states due to the Coulomb interaction promotes  the photoinduced kinetic-energy suppression. 

\begin{figure}[t]
\begin{center}
\includegraphics[width=\columnwidth, clip]{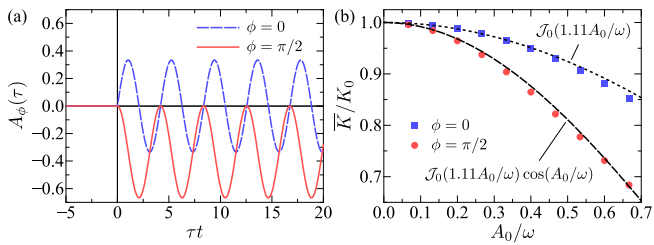}
\end{center}
\caption{(Color online) 
(a) Time profiles of vector potentials $A_{\phi}(\tau)$ at $\phi=0$ and $\pi/2$.
(b) ${\overline K}/K_0$ at $\phi=0$ (filled squares) and $\pi/2$ (filled circles), calculated using  iTEBD method. 
Parameter values are chosen to be $U/t=8, n=0.6854$ and $\omega /t = 1.5$.
The dotted line indicates the least-square fitting of the data at $\phi=0$ and the 
broken line is calculated from the dotted line multiplied by a factor of $\cos(A_0/\omega)$. 
}
\label{fig:phase}
\end{figure}

We show the effects of the phase in the AC field. 
The phase degree of freedom $\phi$ is introduced in the vector potential as 
$A_\phi(\tau)=\theta(\tau)(A_0/\omega) \left [ \sin\left( \omega \tau+\phi\right ) -\sin \phi \right] $. 
When $\phi=0$, $A_{\phi}(\tau)$ reduces to Eq.~(\ref{eq:ac}) with $\tau_{\rm p}=0$, and the oscillation in $A_{\phi}(\tau)$ is symmetric with respect to the origin, as shown in Fig.~\ref{fig:phase}(a). 
On the other hand,  when $\phi=\pi/2$, the oscillation in $A_{\phi}(\tau)$ is asymmetric, but that in the electric field is symmetric. 
In the non-interacting system, we have an exact expression, i.e.,   $\overline{K}/K_0=\mathcal{J}_0(A_0/\omega) \cos[(A_0/\omega) \sin\phi] $, where an apparent cosine factor appears as a result of the phase. 
The phase effect introduced in the Hubbard Hamiltonian is examined using the iTEBD method. 
The results obtained for $U/t=8$ and $n=0.685$ are presented in Fig.~\ref{fig:phase}(b); 
$\overline{K}$ at $\phi=0$ and $\pi/2$ are well fit by $\mathcal{J}_0(c_{\rm K} A_0/\omega)$ and $\mathcal{J}_0(c_{\rm K}A_0/\omega) \cos(A_0/\omega)$, respectively, where $c_{\rm K}=1.11$ in both cases. 
It is concluded that the deviation of $c_{\rm K}$ from one does not depend on the choice of phase, but it is attributable to the intrinsic effects. 

Thus far, we have primarily focused on the results for systems far from the half filling $(n\lsim 0.8)$. 
This is because the analysis accuracy of this case for the Floquet+ED and perturbation methods is limited. 
The short-range magnetic interaction of the order of $t^2/U$ provides an additional energy scale near the half filling, and may play a role in the suppression of $c_{\rm K}$ under the AC field. 
Although the results yielded by the iTEBD method shown in Fig.~\ref{fig:main} are reliable even near the half filling, further analyses are required as future research questions. 
It is likely that the size effect on the ED method, the Floquet states that are neglected during implementation of  the perturbation methods, and the higher-order perturbations are considered in that case. 

The present study reports several  invaluable findings for experimental observations of  the electron motion suppression under an AC field. 
First, for wide $U/t$ range from the weak to strong coupling regimes, and electron density, 
the reduction of $\overline{K}/K_0$ is well scaled by the modified zeroth-order Bessel function for small $A_0/\omega$ at least. 
Second, the suppression is most remarkable at approximately $n=0.8$ and $U/t=10$, and is reduced in the vicinity of the half-filled state. This finding aids appropriate selection of target materials for experimental observations. 
Third, the phase of the AC field does not play an essential role in the suppression; this finding provides valuable information for setting up the light pulse in experiment. 
Theoretical calculations for higher dimensional systems are required for direct comparison with experiment, although the present results via the perturbation method are obtained without assumptions regarding the system dimensions. 
 
In summary, the correlated electron dynamics under an AC field is examined for a one-dimensional Hubbard model. 
Through the analyses using the three complementary methods,  
it is found that the photoinduced kinetic-energy suppression itself is influenced sensitively by the on-site Coulomb interaction as well as the electron density.
The results are interpreted as a combination effect of the electron redistribution via the Coulomb interaction 
and the AC field. 
The present results will be checked directly through systematic experiments involving a series of low-dimensional conducting organic solids as well as cold-atom systems. 

The authors would like to thank M. Naka and S. Iwai for their fruitful discussions. 
This work was supported by MEXT KAKENHI, Grant Numbers 26287070 and 15H02100. 
Some of the numerical calculations were performed using the facilities of the Supercomputer Center, the Institute for Solid State Physics, the University of Tokyo.



\clearpage
\begin{center}
{\large{\bf Supplemental Material for\\``Optical suppression of electron motion in low-dimensional correlated electron system''}}
\end{center}

In this Supplemental Material, we derive the analytical expression of the ``suppression factor'' $c_{\rm K}$ shown in Eq.~(6) in the main text by using the perturbation method. 
In the case of the small $\mathcal{A} \equiv A_0/\omega$, 
this is defined as a coefficient in the time-averaged kinetic energy given as 
\begin{align}
\overline{K}/K_0 = \mathcal{J}_0(c_{\rm K}\mathcal{A}) 
\sim 1 - \frac{1}{4} (c_{\rm K}\mathcal{A})^2 + \mathcal{O}(\mathcal{A}^3) .
\tag{S.1} \label{eq:def}
\end{align}
We adopt the Hubbard model ${\mathcal H}$ in a one-dimensional chain in the main text, although the following formulae  are given in a $d$-dimensional lattice.
%

In the Floquet theory, the $m$th-order Fourier component of the time-dependent Hamiltonian is given by 
\begin{align}
H_m
&= \frac{\omega}{2\pi} \int_0^{2\pi/\omega}d\tau\, e^{im\omega\tau} \mathcal{H}_A \notag \\
&= \delta_{m,0} \mathcal{H}_{\rm int} + \begin{cases}
\mathcal{J}_m(\mathcal{A}) \mathcal{H}_{\rm kin} & (m: \text{even}) ,  \\
\mathcal{J}_m(\mathcal{A}) (-iJ) & (m: \text{odd}),
\end{cases}
\tag{S.2}
\end{align}
where $\mathcal{H}_A$ is the Hubbard Hamiltonian where $A(\tau)$ is taken into account as the Peierls phase. 
We introduce $J = \sum_{\bm{k}\sigma} \sum_{\nu=1}^d (2t\sin k_\nu) c_{\bm{k}\sigma}^\dagger c_{\bm{k}\sigma}$ with the Fourier component of the fermion operator $c_{\bm{k}\sigma} = L^{-d/2} \sum_{j} e^{-i\bm{k}\cdot \bm{R}_j} c_{j\sigma}$.
The Floquet eigen-value equation is given by 
\begin{align}
H^\omega \vert \phi_\alpha \rangle = \varepsilon_\alpha \vert \phi_\alpha \rangle , 
\tag{S.3} \label{eq:eigenvalue-eq}
\end{align}
which corresponds to Eq.~(4) in the main text. 
We introduce a vector space $\{ \vert n \rangle \}_{n \in \mathbb{Z}}$ so that $\vert \phi_\alpha^n \rangle$ and $H_{nm}^\omega$ in Eq.~(4) are represented as $\vert \phi_\alpha^n \rangle = \langle n \vert \phi_\alpha \rangle$ and $H_{nm}^\omega = \langle n \vert H^\omega \vert m \rangle$, respectively.
The AC field is applied to the system in the ground state $\vert 0 \rangle$ of ${\cal H}$ at $\tau=0$. 
An expectation value of an operator $O(\tau)$ at time $\tau$ is given by
\begin{align}
\langle \Psi(\tau) \vert O(\tau) \vert \Psi(\tau) \rangle
&= \sum_{\alpha\beta} \sum_{mn} c_\alpha^\ast c_\beta e^{-i(\varepsilon_\beta - \varepsilon_\alpha) \tau} \notag \\
&\quad \times e^{-i (n-m) \omega \tau} \langle \phi_\alpha^m \vert O(\tau) \vert \phi_\beta^n \rangle,
\tag{S.4} \label{eq:expectationvalue}
\end{align}
where $\Psi(\tau)$ is the wave function at $\tau$, and $c_\alpha=\langle \phi_\alpha^{n=0} \vert 0 \rangle$.

Equation~\eqref{eq:expectationvalue} is evaluated by the perturbation theory. 
The Floquet Hamiltonian $H^\omega$ in Eq.~\eqref{eq:eigenvalue-eq} 
is separated into the unperturbed and perturbed components as
\begin{align}
H^\omega &= V_0 + V_{A} , 
\tag{S.5}
\end{align}
where $V_0$ is defined as $ H^\omega$ at $A_0=0$, and $V_A$ is the remaining part of $H^\omega$. 
The quasienergy and the wave function in the $j$th-order perturbation are given by 
$\varepsilon_\alpha^{(j)}$ and
$\vert \phi_\alpha^{(j)} \rangle$, respectively. 
For the non-degenerate eigenstate of $V_0$, the eigenstate of $H^\omega$ is obtained up to the first order of $V_{A}$ as 
\begin{align}
\vert \phi_\alpha \rangle &= \vert \phi_\alpha^{(1)} \rangle + \mathcal{N} \vert \phi_\alpha^{(0)} \rangle + \mathcal{O}(V_A^2) \notag \\
&\approx \sum_{\beta (\neq \alpha)} \vert \phi_\beta^{(0)} \rangle \frac{ \langle \phi_\beta^{(0)} \vert V_{A} \vert \phi_\alpha^{(0)} \rangle }{ \varepsilon_\alpha^{(0)} - \varepsilon_\beta^{(0)} } + \mathcal{N} \vert \phi_\alpha^{(0)} \rangle, 
\tag{S.6} \label{eq:floquetstate}
\end{align}
where a constant $\mathcal{N}$ is determined by the normalization.
Since the unperturbed Hamiltonian $V_0$ is block diagonal, Eq.~\eqref{eq:eigenvalue-eq} is decomposed into each sector as 
\begin{align}
\left (H_0|_{\mathcal{A}=0} - m\omega \right) \vert \phi_\alpha^{m (0)} \rangle = \varepsilon_\alpha^{(0)} \vert \phi_\alpha^{m (0)} \rangle .
\tag{S.7}
\end{align}
where $H_0|_{\mathcal{A}=0}$ is $H_{m=0}$ at  ${\mathcal A}=0$, 
and is nothing but the Hubbard Hamiltonian ${\cal H}$. 
Thus, the zeroth-order eigenstates $\vert \phi_\alpha^{m (0)} \rangle$ are identified as the eigenstates of ${\mathcal H}$, and the Floquet states in Eq.~\eqref{eq:floquetstate} are represented in terms of the eigenstates $\vert i \rangle$ and eigen-energies $E_i$ of ${\mathcal H}$. 

We assume that the ground state of $\mathcal{H}$ is non-degenerated 
and the resonant transitions do not occur, i.e., $m\omega \neq E_i-E_0$.
In Eq.~(\ref{eq:expectationvalue}), 
one Floquet state $\vert \alpha \rangle$, which connects to the ground state $\vert n=0 \rangle \vert 0 \rangle$ in the limit of ${\mathcal A} \rightarrow 0$, is taken into account. 
This is justified in the small $A/\omega$ region by the calculated results shown in Fig.~3(a) in the main text, in which the weight $|c_\alpha|^2$ for one Floquet state is dominant, and others are much less than one. 

By comparing Eq.~\eqref{eq:def} with the following expression: 
\begin{align}
\overline{K} 
&\approx \sum_{mn} \mathcal{J}_{m-n}(\mathcal{A}) \times
\begin{cases}
\langle \phi_\alpha^m \vert \mathcal{H}_{\rm kin} \vert \phi_\alpha^n \rangle & (m-n:\text{even}) ,  \\
\langle \phi_\alpha^m \vert (-iJ) \vert \phi_\alpha^n \rangle & (m-n:\text{odd}),
\end{cases}
\tag{S.8}
\end{align}
we obtain the analytical expression for $c_{\rm K}$ as follows, 
\begin{widetext}\begin{align}
c_{\rm K}^2 &= 1 + \frac{2}{K_0} \sum_{i (\neq 0)} \frac{ \vert \langle i \vert \mathcal{H}_{\rm kin} \vert 0 \rangle \vert^2 }{ E_0 - E_i } + \sum_{i (\neq 0)} \left[ \frac{\vert \langle i \vert J \vert 0 \rangle \vert^2}{ \left( E_0 - (E_i - \omega) \right)^2 } + \frac{ \vert \langle i \vert J \vert 0 \rangle \vert^2 }{ \left( E_0 - (E_i + \omega) \right)^2 } \right] \notag \\
&\quad - \frac{2}{K_0} \sum_{i (\neq 0)} \left[ \frac{ \vert \langle i \vert J \vert 0 \rangle \vert^2 }{ E_0 - (E_i - \omega) } + \frac{ \vert \langle i \vert J \vert 0 \rangle \vert^2 }{ E_0 - (E_i + \omega) } \right]
- \frac{1}{K_0} \sum_{n=\pm 1} \sum_{\substack{i (\neq 0) \\ j (\neq 0)}} \frac{ \langle 0 \vert J \vert i \rangle \langle i \vert \mathcal{H}_{\rm kin} \vert j \rangle \langle j \vert J \vert 0 \rangle }{ \left( E_0 - (E_i - n\omega) \right) \left( E_0 - (E_j - n \omega) \right) }.
\tag{S.9} \label{eq:ck}
\end{align}
The first two terms are Eq.~(6) in the main text.
In the limits of the low-energy ($\omega \rightarrow 0$) and the weak excitation ($\mathcal{A}\rightarrow 0$), this is reduced to 
\begin{align}
c_{\rm K}^2 = 1 + 2 \sum_{i(\neq 0)} \left[ \frac{ \vert \langle i \vert \mathcal{H}_{\rm kin} \vert 0 \rangle \vert^2 }{ K_0 (E_0-E_i) } - \frac{ 2\vert \langle i \vert J \vert 0 \rangle \vert^2 }{ K_0 (E_0 - E_i) } + \frac{ \vert \langle i \vert J \vert 0 \rangle \vert^2 }{ (E_0-E_i)^2 } \right]
- 2 \sum_{\substack{i (\neq 0) \\ j (\neq 0)}} \frac{ \langle 0 \vert J \vert i \rangle \langle i \vert \mathcal{H}_{\rm kin} \vert j \rangle \langle j \vert J \vert 0 \rangle }{ K_0 (E_0 - E_i) (E_0 - E_j) }.
\tag{S.10}
\end{align}
On the other hand, in the limit of the high-energy excitation ($\omega \rightarrow \infty$),
we have 
\begin{align}
c_{\rm K}^2 = 1 + 2 \sum_{i (\neq 0)} \frac{ \vert \langle i \vert \mathcal{H}_{\rm kin} \vert 0 \rangle \vert^2 }{ K_0 (E_0 - E_i) }.
\tag{S.11}
\end{align}
\end{widetext}


\begin{thebibliography}{99} 

\bibitem{koshihara}
S. Koshihara, and M. Kuwata-Gonokami (eds.),
J. Phys. Soc. Jpn. {\bf 75}, 011001-011008 (2006).

\bibitem{basov}
D. N. Basov,  R. D. Averitt,  D. van der Marel, M. Dressel, and K. Haule,    
Rev. Mod. Phys. {\bf 83}, 471541 (2011).

\bibitem{aoki}
H. Aoki, N. Tsuji, M. Eckstein, M. Kollar, T. Oka, and P. Werner, 
Rev. Mod. Phys. {\bf 86}, 779 (2014).

 
\bibitem{ichikawa} 
H. Ichikawa, S. Nozawa, T. Sato, A. Tomita, K. Ichiyanagi, M. Chollet, L. Guerin, N. Dean, A. Cavalleri, S. Adachi, T. Arima, H. Sawa, Y. Ogimoto, M. Nakamura, R. Tamaki, K. Miyano, and S. Koshihara, Nat. Mater. {\bf 10}, 101 (2011).

\bibitem{stojchevska}
L. Stojchevska, I. Vaskivskyi, T. Mertelj, P. Kusar, D. Svetin, S. Brazovskii, and D. Mihailovic, 
Science {\bf 344}, 177 (2014).

\bibitem{kaiser}
S. Kaiser, S. R. Clark, D. Nicoletti, G. Cotugno, R. I. Tobey, N. Dean, S. Lupi, H. Okamoto, T. Hasegawa, D. Jaksch, and A. Cavalleri, Sci. Rep. {\bf 4,} 3823 (2014).

\bibitem{matsuzaki}
H. Matsuzaki, M. Ohkura, Y. Ishige, Y. Nogami, and H. Okamoto, 
Phys. Rev. B {\bf 91}, 245140 (2015).

\bibitem{zhang}
J. Zhang, X. Tan, M. Liu, S. W. Teitelbaum, K. W. Post, F. Jin, K. A. Nelson, D. N. Basov, W. Wu, and R. D. Averitt, Nat. Mater. {\bf 15}, 956 (2016).

\bibitem{moussa}
N. O. Moussa, G. Molnar, S. Bonhommeau, A. Zwick, S. Mouri, K. Tanaka, J. A. Real, and A. Bousseksou, 
Phys. Rev. Lett. {\bf 94}, 107205 (2005).  
 


\bibitem{tsuji}
N. Tsuji, T. Oka, P. Werner, and H. Aoki,  
Phys. Rev. Lett. {\bf 106}, 236401 (2011). 

\bibitem{mentink}
J. H. Mentink, K. Balzer, and M. Eckstein, 
Nat. Comm. {\bf 6}, 6708 (2015). 


%
\bibitem{dunlap}
D. H. Dunlap and V. M. Kenkre, Phys. Rev. B {\bf 34}, 3625 (1986).
%
\bibitem{grossmann} 
F. Grossmann, T. Dittrich, P. Jung, and P. Hanggi,
Phys. Rev. Lett. {\bf 67}, 516 (1991).
%
\bibitem{kayanuma}
Y. Kayanuma and K. Saito, 
Phys. Rev. A {\bf 77}, 010101(R) (2008).
%

\bibitem{eckardt}
A. Eckardt, C. Weiss, and M. Holthaus, Phys. Rev. Lett. {\bf 95}, 260404 (2005). 

\bibitem{lignier}
H. Lignier, C. Sias, D. Ciampini, Y. Singh, A. Zenesini, O. Morsch, and E. Arimondo,
Phys. Rev. Lett. {\bf 99}, 220403 (2007).

\bibitem{nishioka}
K. Nishioka and K. Yonemitsu,
J. Phys. Soc. Jpn. {\bf 83}, 024706 (2014).
%
\bibitem{yonemitsu}
K. Yonemitsu and K. Nishioka,
J. Phys. Soc. Jpn. {\bf 84}, 054702 (2015).

\bibitem{ishikawa}
T. Ishikawa, Y. Sagae, Y. Naitoh, Y. Kawakami, H. Itoh, K. Yamamoto, K. Yakushi, H. Kishida, T. Sasaki, S. Ishihara, Y. Tanaka, K. Yonemitsu, and S. Iwai, 
Nat. Comm. {\bf 5}, 5528 (2014). 

\bibitem{naitoh}
Y. Naitoh, Y. Kawakami, T. Ishikawa, Y. Sagae, H. Itoh, K. Yamamoto, T. Sasaki, M. Dressel, S. Ishihara, Y. Tanaka, K. Yonemitsu, and S. Iwai, 
Phys. Rev. B {\bf 93}, 165126 (2016).


%
%
\bibitem{vidal}
G. Vidal, 
Phys. Rev. Lett. {\bf 98}, 070201 (2007).
\bibitem{ors}
R. Orus and G. Vidal, 
Phys. Rev. B {\bf 78}, 155117 (2008).
%
\bibitem{takayoshi}
S. Takayoshi, M. Sato, and T. Oka, 
Phys. Rev. B {\bf 90}, 214413 (2014). 
%
\bibitem{ono}
A. Ono, H. Hashimoto, and S. Ishihara, 
Phys. Rev. B {\bf 94}, 115152 (2016). 
%
\bibitem{mcculloch}
I. McCulloch, arXiv:0804.2509. 
%
%
%
\bibitem{shirley}
J. Shirley, Phys. Rev. {\bf 138}, B979 (1965). 
%
\bibitem{sambe}
H. Sambe, Phys. Rev. A {\bf 7}, 2203 (1973). 

\bibitem{sm}
See Supplemental Material at http://link.aps.org/supplemental/
**** for detailed formulations.

\end{thebibliography}
\end{document}